\documentclass{article}
\usepackage{spconf,amsmath,amsfonts,graphicx,url}

\title{Joint Acoustic Echo Cancellation and Speech Dereverberation Using Kalman filters}
\name{Ziteng Wang, Yueyue Na, Biao Tian, Qiang Fu}
\address{Alibaba Group, China}
\begin{document}
%
\maketitle
\begin{abstract}
This paper proposes a joint acoustic echo cancellation (AEC) and speech dereverberation (DR) algorithm in the short-time Fourier transform domain. The reverberant microphone signals are described using an auto-regressive (AR) model. The AR coefficients and the loudspeaker-to-microphone acoustic transfer functions (ATFs) are considered time-varying and are modeled simultaneously using a first-order Markov process. This leads to a solution where these parameters can be optimally estimated using Kalman filters. It is shown that the proposed algorithm outperforms vanilla solutions that solve AEC and DR sequentially and one state-of-the-art joint DRAEC algorithm based on semi-blind source separation, in terms of both speech quality and echo reduction performance.
\end{abstract}
\section{Introduction}

Acoustic echo cancellation (AEC) and speech dereverberation (DR) play important roles in speech processing applications like real-time communications and automatic speech recognition. AEC aims to remove echos introduced by the playback signals. DR aims to reduce reverberation and restore direct path sounds, optionally retaining early reflections. Adaptive filters are widely used in the two tasks, such as the normalized least means square (NLMS) filter~\cite{shynk1992frequency}, the recursive least square (RLS) filter, and the weighted prediction error (WPE) algorithm~\cite{yoshioka2012generalization}. Kalman filters have also been separately applied in AEC~\cite{enzner2006frequency,paleologu2013study,yang2017frequency,yang2022low} and DR~\cite{schwartz2014online,braun2016online,braun2018linear,lemercier2022neural}.

From the view of adaptive filtering, AEC and DR share many similarities in their solving process. Both tasks are related to estimation of the room impulse responses (RIRs), one relating a source position to the microphone position and the other relating a loudspeaker position to the microphone position.
Joint algorithms that deal with AEC and DR at the same time have shown performance gains over separate partial algorithms~\cite{takeda2009ica,takeda2012efficient,togami2014simultaneous,carbajal2020joint}.
Takeda et al.~\cite{takeda2009ica} achieve blind dereverberation and echo cancellation by applying a frequency domain independent component analysis (ICA) model. 
It is assumed that direct sound frame is independent of late reverberation and playback signals, which is approximately true under multiple input/output inverse filtering theorem (MINT) conditions.
Togami and Kawaguchi \cite{togami2014simultaneous} combine acoustic echo reduction, speech dereverberation and noise reduction by assuming a time-varying local Gaussian model of the microphone input signal. The linear filters are optimized under a unified likelihood function that directly reflects the eventual speech enhancement performance.
Cohen et al.~\cite{cohen2021online} propose a Kalman-EM method based on a moving average model for speech dereverberation, and the Kalman filter is adopted to estimate the clean signal after echo cancellation in the E-step.

The reverberant signals are often described by an auto-regressive (AR) model in the short time Fourier transform (STFT) domain.
Our previous works~\cite{wang2021weighted,wang2020semi} reformulate AEC and DR from the semi-blind source separation perspective, respectively. A RLS based joint DRAEC algorithm is further derived in~\cite{na21_interspeech}. 
Considering RLS can be seen a special case of Kalman filter~\cite{sayed1994state}, we further propose in this paper a joint DRAEC algorithm using Kalman filters.
The AR model coefficients and the loudspeaker-to-microphone acoustic transfer functions (ATFs) are considered time-varying and are modeled simultaneously using a first-order Markov process.
By minimizing a unified mean squared error loss function, a novel joint DRAEC algorithm is derived. The joint algorithm not only outperforms its RLS counterpart but also outperforms cascaded alternatives that use Kalman based AEC and Kalman based DR sequentially. 


\section{Signal model}

We consider a multi-channel convolutive mixture in the short-time Fourier transform (STFT) domain. A sensor array of $M$ microphones captures signals from source $S(t,f)$ and signal $X(t,f)$ played by a loudspeaker, with $t$ and $f$ the time index and the frequency index, respectively. 
The $m$th microphone signal in the $f$th band is given by:
\begin{align}\label{eq:signalmodel}
\nonumber	Y_m(t) = & \sum_{l=0}^{\infty}A_{m,l}(t)S(t-l) \\ 
	& + \sum_{l=0}^{\infty}B_{m,l}(t) X(t-l) + V_m(t)
\end{align}
where $A_{m,l}$ denotes the source-to-microphone transfer function, $B_{m,l}$ denotes the loudspeaker-to-microphone transfer function, and $V_m$ denotes the background noise. 
The signal can be approximated by an auto-regressive model~\cite{togami2014simultaneous} as:
\begin{align}\label{eq:armodel}
	\nonumber	Y_m(t) = & S_m(t) + \sum_{n=1}^{M}\sum_{l=0}^{L_Y-1}C_{m,n,l}Y_n(t-\Delta-l)\\ 
	& + \sum_{l=0}^{L_X-1}B_{m,l}(t) X(t-l) + \tilde{V}_m(t)
\end{align}
where $S_m$ is the direct path sound and early reflections, $\Delta$ marks the boundary between early reflections and late reverberation, $C_{m,n,l}$ denotes the multichannel auto-regressive coefficients, and $\tilde{V}$ contains the modeling error.

The first term in the right-hand side of (\ref{eq:armodel}) is defined as the target to be recovered:
\begin{equation}\label{eq:filtering}
	S_m(t) = Y_m(t) - {\bf w}^H_m(t){\bf z}(t)
\end{equation}
where 
\begin{align}\label{eq:filterw}
	\nonumber &	{\bf w}_m  = [
	\underset{{\bf w}_\text{AEC}}{\underbrace{ B^*_{m,0}, ..., B^*_{m,L_X-1}, }}\\ 
	& \underset{{\bf w}_\text{DR}}{\underbrace{ C^*_{m,1,0}, ..., C^*_{m,1,L_Y-1}, ..., C^*_{m,M,0}, ..., C^*_{m,M,L_Y-1} }} ]^T 
\end{align}
is a unified vector consisting of the AEC related filters of length $L_X$ and the DR related filters of length $ML_Y$, and
\begin{align}
\nonumber	{\bf z}(t) = [& X(t), ..., X(t-L_X+1), \\ 
\nonumber	& Y_1(t-\Delta), ..., Y_1(t-\Delta-L_Y+1), \\
\nonumber	& \quad ..., \\
	& Y_M(t-\Delta), ..., Y_M(t-\Delta-L_Y+1)]^T
\end{align}
is a concatenation of the playback signals and the time-delayed microphone observations. $(\cdot)^*$ denotes complex conjugate, $(\cdot)^H$ denotes Hermitian transpose and $(\cdot)^T$ denotes transpose.

An optimal estimate of the filter coefficients (\ref{eq:filterw}) can be obtained by minimizing the squared error loss function:
\begin{equation}
	J_{\bf{w}} = \min \sum_{m=1}^{M}E\{|{\bf w}_m(t)-\hat{\bf w}_m(t)|^2\}
\end{equation}
where $\hat{(\cdot)}$ denotes variable estimate. We assume the following algorithm is performed for each microphone independently, and the subscript $m$ is omitted for brevity.

\begin{figure}[th]
	\centering
	\centerline{\includegraphics[width=0.85\columnwidth]{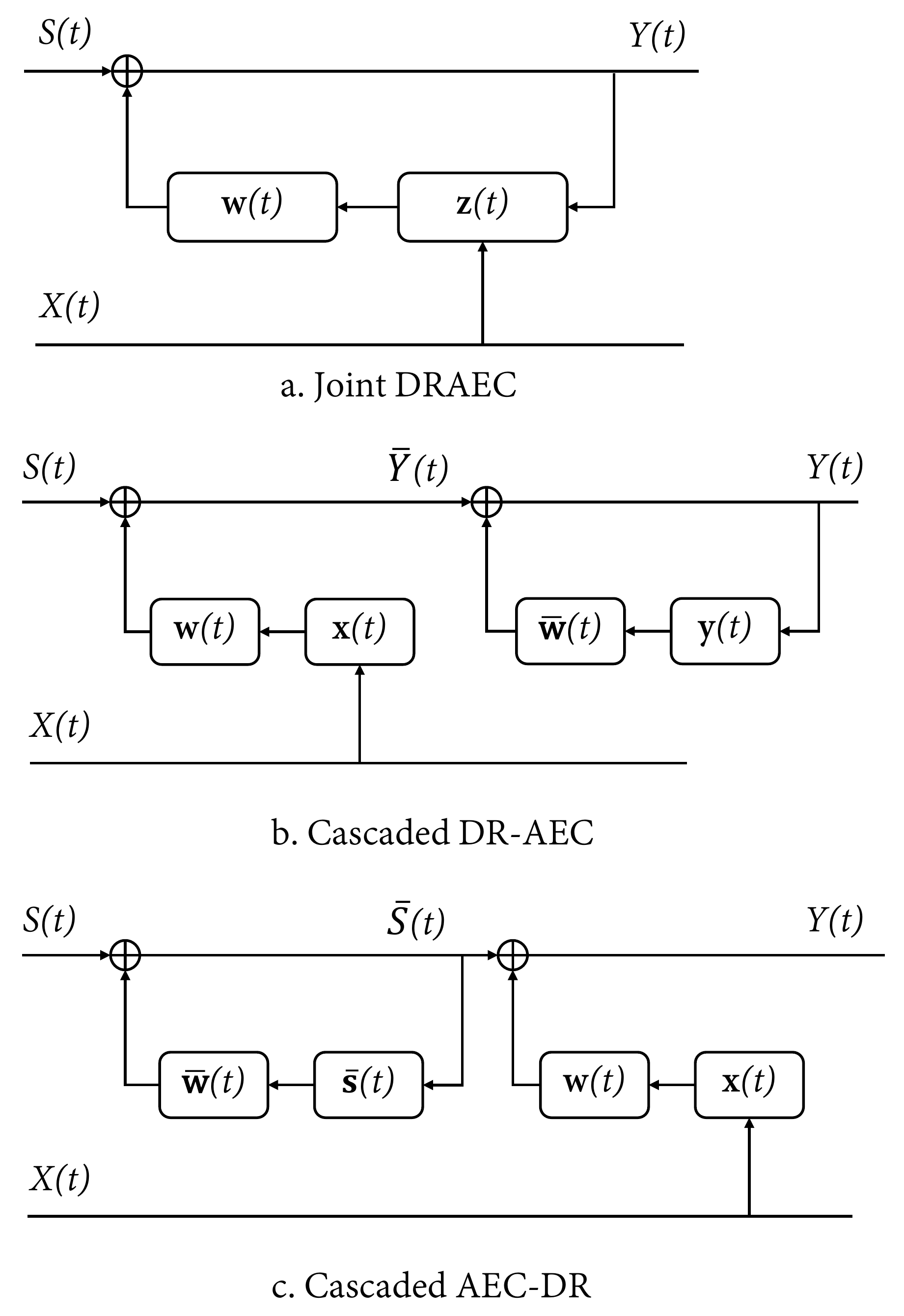}}
	\caption{Source models of joint DRAEC and cascaded DR-AEC and AEC-DR.}
	\label{fig1}
\end{figure}

\section{The proposed algorithm}

Both the source-to-microphone and loudspeaker-to-microphone transfer functions are time-varying in real acoustic scenarios, therefore the filter states $\{B_l, C_{n,l}\}$ are also considered to be time-varying. This is described by a first-order Markov process:
\begin{equation}
	{\bf w}(t) = {\bf}A(t) *{\bf w}(t-1) + {\bf u}(t) 
\end{equation}
where ${A}(t)$ is the state transition parameter and ${\bf u}(t) \sim \mathcal{N}({\bf 0}, {\bf \Phi}_{\bf u}(t))$ describes the process noise that follows a zero-mean complex Gaussian distribution.

\subsection{Kalman based DRAEC}

Given the above formulation, the well-known Kalman filter\cite{kalman1960new} is applied to estimate ${\bf w}(t)$. We denote ${\bf \Phi}$ as the state vector error covariance matrix:
\begin{equation}
	{\bf \Phi}(t) = E\{[{\bf w}(t)-\hat{\bf w}(t)][{\bf w}(t)-\hat{\bf w}(t)]^H\}
\end{equation}
The filter update equations are given by:
\begin{align}
	 S(t|t-1) &= Y(t) - {\bf w}^H(t|t-1){\bf z}(t) \\
	{\hat{\bf w}}(t) &= {\bf w}(t|t-1)+{\bf k}(t)S^*(t|t-1) \\
	{\bf \Phi}(t) &=[{\bf I} - {\bf k}(t){\bf z}^H(t)]{\bf \Phi}(t|t-1)
\end{align}
where the Kalman gain
\begin{equation}
	{\bf k}(t) = \frac{{\bf \Phi}(t|t-1){\bf z}(t)}{
		\phi_S(t)+{\bf z}^H(t){\bf \Phi}(t|t-1){\bf z}(t)},
\end{equation}
$\phi_S(t)=E\{S(t)S^*(t)\}$ denotes power spectral density of the desired source and $\bf I$ is a unit matrix of proper size. The next time step predict equations are given by:
\begin{align}
	& {\bf w}(t+1|t)= A(t){\hat{\bf w}}(t) \\
	& {\bf \Phi}(t+1|t) = A^2(t) {\bf \Phi}(t)+{\bf \Phi}_{\bf u}(t)
\end{align}
The target source is eventually recovered by:
\begin{equation}\label{eq:filtering}
	\hat{S}(t) = Y(t) - {\hat{\bf w}}^H(t){\bf z}(t)
\end{equation}

\subsection{Cascaded solutions}

Fig.~\ref{fig1} depicts source models of joint DRAEC and alternative cascaded DR-AEC and cascaded AEC-DR, where the last two are named by their solving orders. 
For comparison, we formulate the solution of AEC-DR as:
\begin{align}
	\nonumber & \bar{S}(t) = Y(t) - {\bf w}_{\text{AEC}}^H(t){\bf x}(t) \\
	& S(t) = \bar{S}(t) - \bar{{\bf w}}_{\text{DR}}(t)\bar{\bf s}(t)
\end{align}
where ${\bf x}(t)$ is a vector of the playback signals, and $\bar{\bf s}(t)$ is a vector of the time delayed signals $\{\bar{S}_m(t-\Delta-l)\}$ after AEC. 
Note that the AR coefficients $\bar{{\bf w}}_{\text{DR}}$ are different from ${\bf w}_{\text{DR}}$ as defined in (\ref{eq:filterw}), because the involved signals are different.
The DR filter is thus susceptible to performance of the AEC filter.
The loss function in this case is given by: 
\begin{align}
	\nonumber J_{{\bf w}} = & \min E\{|{\bf w}_{\text{AEC}}(t)-\hat{\bf w}_{\text{AEC}}(t)|^2\}\\
	&  + \min E\{|\bar{\bf w}_{\text{DR}}(t)-{\hat{\bar{\bf w}}}_{\text{DR}}(t)|^2\}
\end{align}
and it can be optimized by performing Kalman filter based AEC and Kalman filter based DR sequentially.

\subsection{Parameter estimation}

The Kalman filter requires suitable estimators for ${\bf \Phi}_{\bf u}(t)$ and $\phi_S(t)$. Similarly as in \cite{braun2016online}, we use ${\bf \Phi}_{\bf u}(t) = \phi_{\bf u}(t) {\bf I}$, assuming the elements in ${\bf w}(t)$ uncorrelated and identically distributed. The variance parameter
\begin{equation}
	\phi_{\bf u}(t) = \frac{1}{L}E\{|{\hat{\bf w}}(t) - {\hat{\bf w}}(t-1)|^2\} + \eta
\end{equation}
is estimated by the change of filter coefficients over time, where $L$ denotes the filter length and $\eta$ is a small positive number to retain the tracking ability when the acoustic environment changes. 

A maximum likelihood estimation of $\phi_S$ is given by:
\begin{equation}
	\hat{\phi_S}(t) = \alpha \phi(t-1) + (1-\alpha)S(t|t-1)S^*(t|t-1)
\end{equation}
and 
\begin{equation}
	\phi(t) = \alpha \phi(t-1) + (1-\alpha)\hat{S}(t)\hat{S}^*(t)
\end{equation}
where $\alpha$ is a recursive smoothing factor.

For initialization, we use ${\bf w}(0) = {\bf 0}$ and ${\bf \Phi}(0)={\bf I}$.

\section{Experiments}

\begin{figure}[th]
\centering
\centerline{\includegraphics[width=1.05\columnwidth]{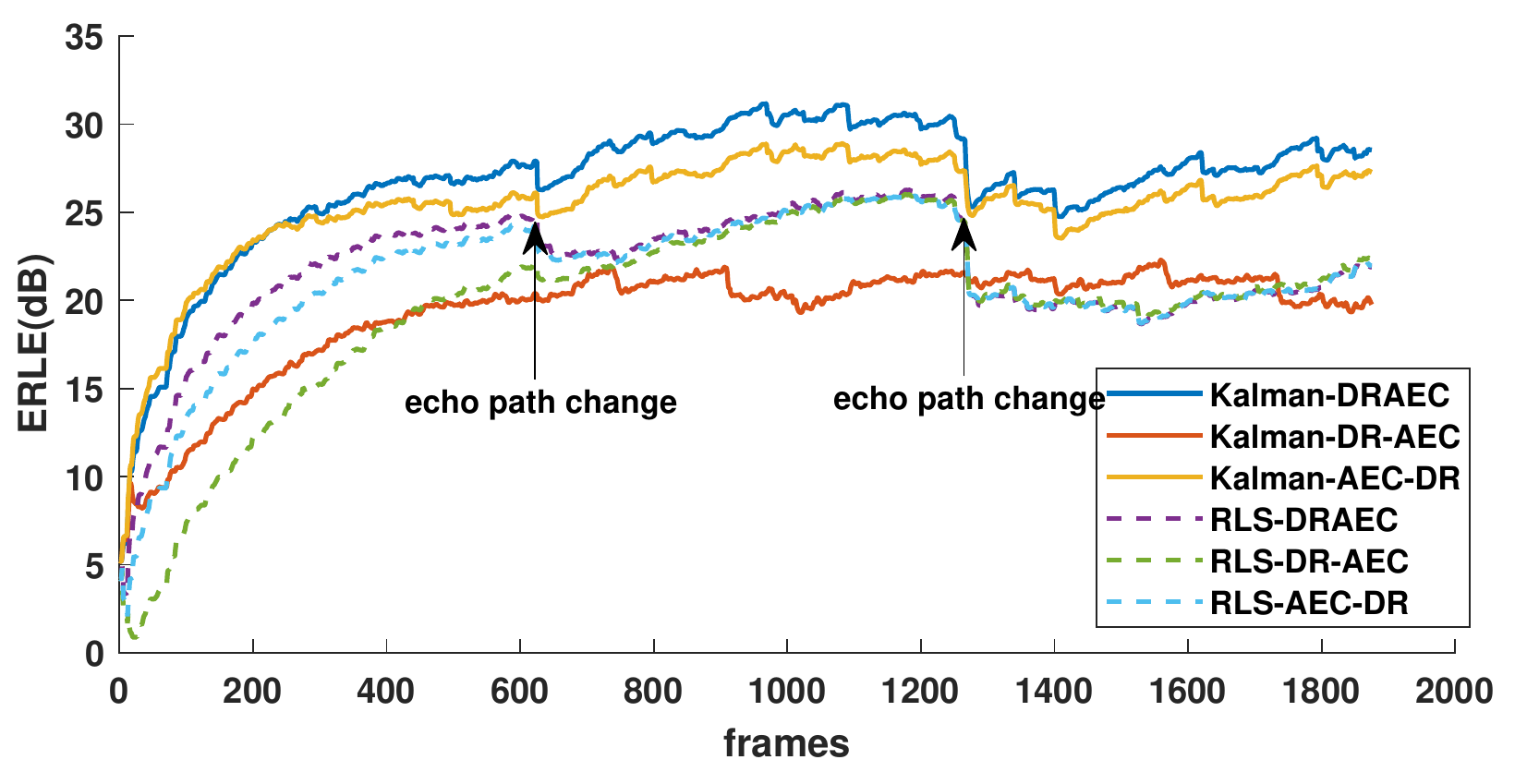}}
\caption{ERLE when echo path changes. Best viewed in color.}
\label{fig2}
\end{figure}

The experiments are conducted in echoic, echoic \& reverberant, and echoic \& reverberant \& noisy environments. Reverberation only is not considered because the source models as in shown Fig.~\ref{fig1} would degenerate to the same one.

The sampling frequency is 16 kHz. We use a frame size of 32 ms, 50\% overlap between frames and a STFT size of 1024 points. The AEC filter length is set to $L_X=5$ and the DR filter length is $L_Y=5$ with delay $\Delta=2$.
We use $A=1$, $\eta=1e^{-4}$ and $\alpha=0.8$ in the Kalman filters.  The experimental setup mainly follows that as in~\cite{na21_interspeech}. The proposed algorithm is compared with its RLS based variants, the implementations of which are open sourced\footnote{https://github.com/nay0648/unified2021}. 


\subsection{Echo}

We first consider the task of single talk echo cancellation. The echo signals are recorded using a smart speaker with $M=2$ microphones and one loudspeaker. In Fig.~\ref{fig2}, echo return loss enhancement (ERLE), defined as ratio of the input signal power to the output signal power, is investigated. Echo path change is simulated by concatenating two different test files. Kalman-DRAEC achieves the highest steady state performance of 31.15 dB ERLE, surpassing the second-best Kalman-AEC-DR by 2.30dB and all the RLS variants. The joint DRAEC algorithms outperforms cascaded alternatives, DR-AEC and AEC-DR. 
One reason is the microphone signal contains an echoic copy of the playback sounds, and filtering on microphone signals helps echo reduction especially when nonlinear echo exists.

\begin{table*}[th]
	\caption{PESQ scores in double-talk periods. In the case of no reverb, clean speech is used as the reference.}
	\label{tab1}
	\centering
	\begin{tabular}{l|ccc|ccc|ccc}
		\hline
		RT60	& \multicolumn{3}{c}{no reverb} & \multicolumn{3}{c}{0.3s}  & \multicolumn{3}{c}{0.6s} \\
		SER & 0dB & -10dB & -20dB & 0dB & -10dB & -20dB & 0dB & -10dB & -20dB \\ \hline
		\text{Orig}
		& 1.06 & 1.07 & 1.04 & 1.14 & 1.07 & 1.07 & 1.11 & 1.08 & 1.07  \\ \hline
		\text{Kalman-DRAEC}
		& 1.54 & \bf 1.24 & \bf 1.13 & 1.92 &\bf 1.40 &\bf 1.11 & 1.57 &\bf 1.32 &\bf 1.10  \\ \hline
		\text{Kalman-DR-AEC}
		& 1.23 & 1.06 & 1.09 & 1.55 & 1.14 & 1.08 & 1.35 & 1.12 & 1.09  \\ \hline
		\text{Kalman-AEC-DR}
		& \bf 1.58 & 1.23 & 1.11 & \bf 1.96 & 1.39 & 1.10 &\bf 1.59 & 1.32 & 1.10 \\ \hline
		\text{RLS-DRAEC}
		& 1.40 & 1.19 & 1.04 & 1.74 & 1.34 & 1.08 & 1.48 & 1.28 & 1.09  \\ \hline
		\text{RLS-DR-AEC}
		& 1.24 & 1.12 & 1.04 & 1.60 & 1.23 & 1.07 & 1.38 & 1.21 & 1.08 \\ \hline
		\text{RLS-AEC-DR}
		& 1.39 & 1.18 & 1.04 & 1.74 & 1.33 & 1.08 & 1.48 & 1.27 & 1.09  \\
		\hline
	\end{tabular}
\end{table*}

Double talk utterances are simulated by adding clean speech to the echo signals at signal-to-echo (SER) ratios of 0 dB, -10 dB and -20 dB. The perceptual evaluation of speech quality (PESQ) scores are evaluated and reported in the first category in Table~\ref{tab1}. Kalman-DRAEC achieves overall highest scores when SER=-10 dB and SER=-20 dB, and scores comparable to Kalman-AEC-DR (1.54 vs 1.58) in SER=0 dB.

\subsection{Echo \& Reverb}

Taking reverberation into account, clean speech is first convolved with room impulse responses before adding with echo signals. The impulse responses are generated in random-sized rooms using the Image method~\cite{allen1979image}. The direct sound with early reverberation (50ms) is used as reference. Double talk PESQ scores are reported in the right parts of Table~\ref{tab1}. The trend is the same as in the echoic environments. Kalman filter based algorithms perform overall better than the RLS based variants. Performing DR before AEC is not recommended because the source model as in Fig.~\ref{fig2}(b) mixes up the source signal and the echo signal.
High reverberation (0.6s) and high echo level (SER=-20dB) is challenging for all the algorithms.

\begin{table}[th]
	\caption{SDR (dB) improvements with respect to input.}
	\label{tab2}
	\centering
	\begin{tabular}{lcccc}
		\hline
		SER	& \multicolumn{2}{c}{0dB} & \multicolumn{2}{c}{-10dB} \\
		RT60	& 0.3s & 0.6s  & 0.3s & 0.6s  \\ \hline
		\text{Kalman-DRAEC}
		& 9.27& 7.68& 16.49& 13.75  \\ \hline
		\text{Kalman-DR-AEC}
		& 8.90& 7.41& 14.75& 12.25  \\ \hline
		\text{Kalman-AEC-DR}
		& 9.37&\bf 7.78& 16.57&\bf 13.77  \\ \hline
		\text{RLS-DRAEC}
		&\bf 9.69& 7.50&\bf 16.82& 13.54  \\ \hline
		\text{RLS-DR-AEC}
		& 9.51& 7.32& 16.05& 12.74  \\ \hline
		\text{RLS-AEC-DR}
		& 9.63& 7.43& 16.76& 13.40  \\
		\hline
	\end{tabular}
\end{table}

\begin{table}[th]
	\caption{SIER (dB) improvements with respect to input.}
	\label{tab3}
	\centering
	\begin{tabular}{lcccc}
		\hline
		SER	& \multicolumn{2}{c}{0dB} & \multicolumn{2}{c}{-10dB} \\
		RT60	& 0.3s & 0.6s  & 0.3s & 0.6s  \\ \hline
		\text{Kalman-DRAEC}
		&\bf 11.54&\bf 9.57&\bf 12.96&\bf 10.02  \\ \hline
		\text{Kalman-DR-AEC}
		& 9.48& 7.85& 8.48 & 6.72  \\ \hline
		\text{Kalman-AEC-DR}
		& 11.36& 9.38& 12.46& 9.22  \\ \hline
		\text{RLS-DRAEC}
		& 11.12& 9.25& 12.09& 8.93  \\ \hline
		\text{RLS-DR-AEC}
		& 10.71& 8.71& 10.40& 6.87  \\ \hline
		\text{RLS-AEC-DR}
		& 10.95& 9.04& 11.90& 8.35  \\
		\hline
	\end{tabular}
\end{table}

\subsection{Echo \& Reverb \& Noise}

The proposed algorithm is also evaluated in complex scenarios where interference and noise coexist. Interfering signals are added at signal-to-interference ratio (SIR) of 0~dB.
Signal-to-distortion ratio (SDR)~\cite{vincent2006performance} and
a non-instructive metrics, namely microphone signal to interference-plus-echo ratio (SIER), are respectively reported in Table~\ref{tab2} and Table~\ref{tab3}. SDR measures the overall quality of the processed speech. SIER measures the non-target reduction performance.

Kalman-DRAEC is advantages in SIER and scores highest in all the test cases, nevertheless, at a cost of more speech distortion as shown in the SDR scores. Based on the previous results, the advantage of DRAEC mainly comes from introducing more echo reduction.

\section{Conclusion}

This paper introduces a joint DRAEC algorithm using Kalman filters that performs echo cancellation and speech dereverberation at the same time. The joint algorithm is derived from a unified mean squared error loss function and outperforms cascaded DR-AEC and AEC-DR alternatives. The Kalman based algorithms also outperforms their RLS based variants because the acoustic environment change is explicitly modeled by a first-order Markov model.

\bibliographystyle{IEEEtran}
\bibliography{refs}

\end{document}